# Frequency-Dependent Breakdown of Litz-Wire Insulation Under Repetitive Bipolar Square-Wave Voltages

Md Asifur Rahman, *Graduate Student Member, IEEE*, Farzana Islam, *Graduate Student Member, IEEE,* and Mona Ghassemi, *Senior Member, IEEE*

***Abstract*—This paper investigates the dielectric breakdown behavior of Litz wire insulation subjected to repetitive fast-edge bipolar square-wave voltages representative of wide-bandgap (WBG) power converter stress. Two conductor sizes, 14 AWG and 16 AWG, were tested in a twisted-pair configuration over frequencies ranging from 1 kHz to 30 kHz using a controlled high-voltage pulse generator with 200 ns rise time. The breakdown voltages obtained at each frequency were statistically analyzed using the weighted two-parameter Weibull method in accordance with IEEE Std 930-2004. Between 1 and 30 kHz, the mean breakdown voltage decreases by approximately 10.7% for the 14 AWG construction and 20.0% for the 16 AWG construction. The 14 AWG specimens exhibit a modest increase between 1 and 5 kHz followed by a decline, whereas the 16 AWG mean decreases across the tested frequencies. The Weibull shape parameter varies nonmonotonically with frequency, characterizing differences in the dispersion of the measured breakdown voltages. These findings provide construction-specific breakdown data relevant to Litz-wire insulation assessment in high-frequency power electronic applications. Possible contributions from charge accumulation, localized discharge activity, and dielectric heating are discussed, but were not directly verified in this study.**

## I. Introduction

THE rapid advancement of high-frequency power electronic systems, driven by the adoption of wide-bandgap (WBG) and ultra-wide-bandgap (UWBG) semiconductor devices, has significantly increased the electrical stress experienced by insulation systems. These devices enable converters to operate at higher switching frequencies and produce fast voltage transitions, improving system efficiency and power density. However, the resulting steep voltage gradients and repetitive square-wave excitations impose severe dielectric stress on motor and transformer windings [1]–[4]. The mechanisms of partial discharge (PD) initiation and dielectric breakdown under such fast-rising square-wave voltages are not yet fully understood, particularly in conductors with complex geometries such as Litz wire.

The authors are with the Department of Electrical and Computer Engineering, The University of Texas at Dallas, Richardson, TX 75080 USA (e-mail: mona.ghassemi@ utdallas.edu).

Litz wire, composed of multiple individually insulated fine strands twisted together, is widely used in high-frequency magnetics to reduce AC resistance due to skin and proximity effects [5]. However, the electrical insulation of Litz wire, typically consisting of thin enamel coatings, was originally optimized for low or moderate frequency sinusoidal excitation. These converters, operating at frequencies that may exceed hundreds of kilohertz and voltage rise times below 100 ns, subject the insulation system of Litz wire to repetitive, high-slew-rate square-wave pulses. These fast-edge pulses significantly raise the electric field stress across turn-to-turn and strand-to-strand interfaces, often triggering partial discharges and progressive insulation erosion. Unlike traditional power-frequency conditions, where breakdown is typically thermally dominated, high-frequency inverter excitation introduces a hybrid failure mechanism involving both electrical overstress and dielectric heating [6], [7].

Previous studies have characterized breakdown and PD behavior in conventional random-wound and tape-insulated stator windings, but relatively few have investigated these phenomena in twisted-pair or Litz wire geometries [8]–[10]. For example, [11] and [12] reported reductions in breakdown voltage and PD inception thresholds under repetitive high-frequency PWM excitations. Recent studies have also employed plasma-based diagnostics to investigate partial-discharge characteristics under fast repetitive pulses, confirming that high-frequency excitation can trigger transient plasma channels and localized ionization in polymer insulation [11]. Their findings confirm that both higher frequencies and steeper pulse edges intensify local field stress and accelerate dielectric aging. Notably, twisted-pair configurations are particularly vulnerable due to field concentration at points of mechanical contact and strand misalignment [13], [14]. However, these studies often employed limited frequencies (typically <100 kHz) and moderate rise times (>200 ns), which do not fully represent the voltage stress conditions encountered in WBG-based converter systems, where rise times may fall below 50 ns.

Under repetitive fast pulses, transient ionization and charge injection can lead to localized plasma microchannels within the enamel coating, similar to those observed in pulsed dielectric breakdown of polymers. These short-lived discharge events may not sustain continuous current but can progressively damage insulation through cumulative plasma-induced erosion.

Moreover, while Weibull statistical analysis has been widely used to describe breakdown behavior, especially for estimating characteristic strength (scale parameter $\alpha$) and predictability (shape parameter β) [15], there remains a need to investigate whether alternate statistical or physical modeling techniques (e.g., log-normal distributions, lifetime testing, and PD pattern tracking) can better capture insulation degradation under repetitive, fast-edge stress.

Qualification of converter-fed motor-winding insulation is addressed by IEC 60034-18-41 for partial-discharge-free Type I systems and IEC 60034-18-42 for partial-discharge-resistant Type II systems [16], [17]. Yet, these specifications primarily target motor windings and do not comprehensively address high-frequency, repetitive square-wave excitation in Litz conductors used in magnetics. Our previous study [18] investigated the frequency-dependent breakdown voltage of a commercial 14 AWG Litz-wire construction under square-wave excitation over 1–30 kHz. The present study compares 14 AWG and 16 AWG commercial constructions under identical bipolar square-wave excitation with a 200 ns rise time, quantifying their differing frequency sensitivities and breakdown-voltage distributions.

Our previous studies have shown that insulation lifetime under repetitive bipolar square-wave excitation need not decrease monotonically with increasing frequency [19]−[22]. In particular, tests of Type I motor-winding insulation at frequencies up to 200 kHz demonstrated increased time to breakdown beyond a critical frequency [19]. Although these fixed-voltage lifetime measurements differ from ramp-to-breakdown voltage measurements, they illustrate the limits of extrapolating frequency trends across operating regimes and support construction-specific experimental assessment.

Published studies of conventional magnet-wire twisted pairs provide important evidence of frequency-dependent dielectric degradation under repetitive voltage excitation. However, these findings do not directly establish the quantitative breakdown characteristics of multistrand Litz-wire bundles, whose arrangements of individually insulated strands, bundle interfaces, and contact regions differ from conventional single-conductor specimens. This study experimentally compares two commercial Litz-wire constructions, 14 AWG and 16 AWG, under identical bipolar square-wave excitation with a 200 ns rise time at five frequencies between 1 and 30 kHz. The contribution is the quantitative characterization of their frequency-dependent breakdown voltages and the comparison of their breakdown-voltage distributions using weighted two-parameter Weibull analysis. The results establish the response of the tested constructions within the specified experimental conditions; they do not establish a universal relationship between conductor gauge, frequency, and dielectric strength.

## II. Experimental Setup and Methodology

### *A. Sample Preparation*

The Litz wire specimens used in this study were prepared to investigate the dielectric breakdown characteristics of stranded conductors under high-frequency voltage stress. Litz wire, composed of multiple individually insulated fine strands twisted together, is widely employed in high-frequency magnetics and power converters to mitigate skin and proximity effects. For the present investigation, two commonly used conductor sizes, 14 AWG and 16 AWG were selected to represent typical wire geometries used in medium-voltage DC (MVDC) systems, high-frequency converters, and power transformer windings. The 14 AWG wire consisted of 260 strands of 38 AWG magnet wire, while the 16 AWG wire contained 162 strands of identical strand size. Both samples utilized solderable polyurethane with a polyamide overcoat (Class F, 155°C) as the insulation material, with overall diameters of 0.090 inch and 0.071 inch, respectively. The principal specifications of the test wires are summarized in Tables I and II. A significant challenge in preparing the samples was insulation removal at the wire ends to form reliable electrical terminations without damaging the fine filaments. Mechanical stripping often fractured the strands, and chemical stripping produced hazardous residues. To mitigate these issues, a thermal stripping method was adopted: the wire ends were immersed in a solder pot containing 99.99% pure tin maintained at 232°C (450°F). The heat dissolved the insulation while simultaneously tinning the strands for subsequent connection. Two insulated Litz-wire lengths were twisted together to form the twisted-pair (TP) specimen. Insulation was removed only at the ends used for electrical termination; the insulation remained intact along the twisted test region. The actual Litz-wound coils may experience additional influences, including mechanical compression, resin impregnation, and thermal cycling. These factors can modify void distribution, permittivity, and stress uniformity. The TP configuration provides a simplified specimen geometry for comparing the two commercial Litz-wire constructions under controlled excitation. It does not isolate intrinsic strand-level properties or reproduce the additional mechanical, impregnation, and thermal conditions of a complete winding. A representative TP section is shown in Fig. 1.

TABLE I
Specifications of 14 AWG Litz wire

| Parameter | Specification |
|---|---|
| Temperature Class | 155°C (311°F), Class F |
| Wire gauge | 14 AWG |
| NEMA Grade | MW 80-C |
| Stranding | 5 bundles of 52 strands of 38 AWG |
| Insulation Material | Solderable Polyurethane with Polyamide Overcoat |
| Overall Diameter | 0.090 inch |

TABLE II
SPECIFICATIONS OF 16 AWG LITZ WIRE

| Parameter | Specification |
|---|---|
| Temperature Class | 155°C (311°F), Class F |
| Wire gauge | 16 AWG |
| NEMA Grade | MW 80-C |
| Stranding | 3 bundles of 54 strands of 38 AWG |
| Insulation Material | Solderable Polyurethane with Polyamide Overcoat |
| Overall Diameter | 0.071 inch |

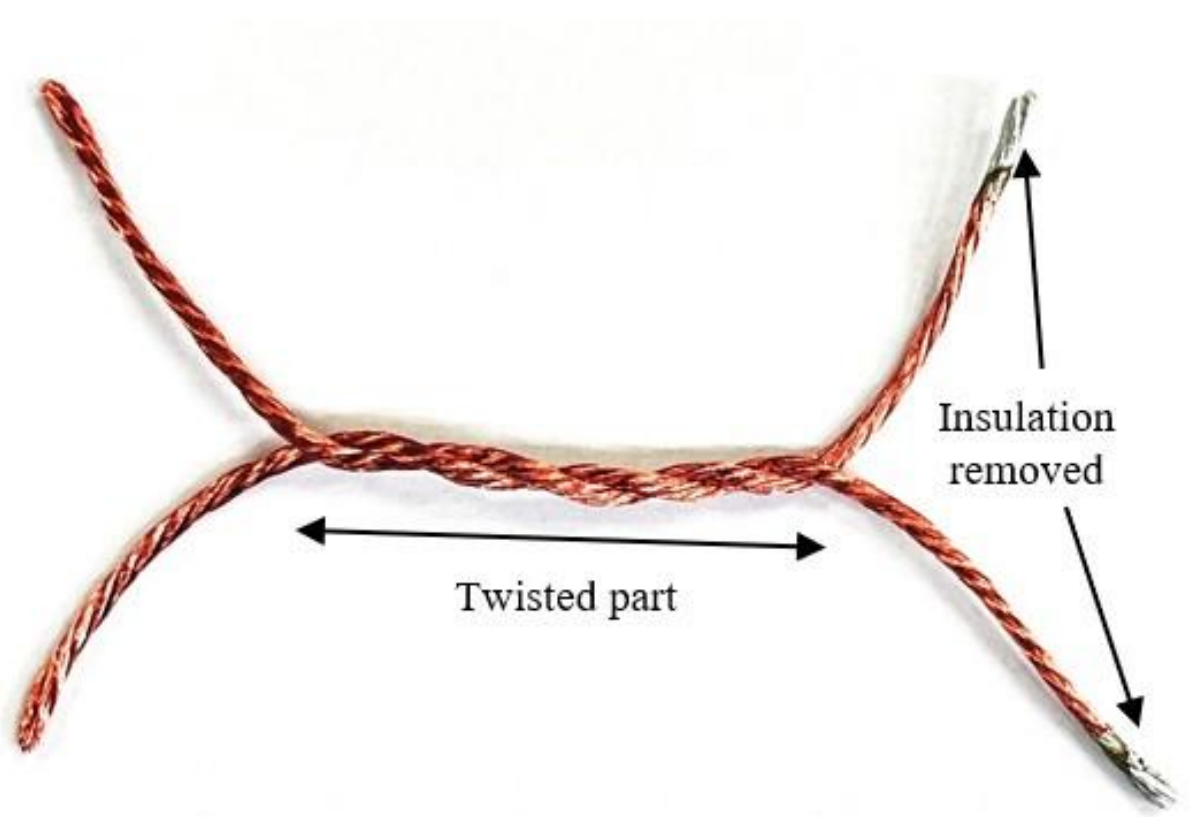

Fig. 1. A TP Litz wire Sample.

Both constructions have the same nominal individual strand gauge, insulation material, and single-build insulation category. Actual enamel thickness and strand packing density were not measured, and twist pitch was not independently varied. No single-strand magnet-wire control group was included. The experiment therefore compares the two commercial constructions as complete specimens and does not isolate the effects of individual construction parameters.

### *B. Experimental Setup*

A high-frequency, high-voltage bipolar pulse generator was employed to apply square-wave voltage stress to the twisted pair (TP) Litz wire samples. The generator could produce both unipolar (positive or negative) and bipolar square voltage pulses up to ±6 kV, with frequency tunability ranging from 0 to 500 kHz. A rise time of 200 ns was selected to replicate the fast voltage transients typically produced by WBG devices such as SiC and GaN switches in high-frequency power converters. It provided precise control over critical waveform parameters such as rise time, frequency, and voltage amplitude, enabling fine-tuned insulation stress testing. Real-time monitoring of the output waveform was performed via a digital oscilloscope connected to the generator's front panel. To maintain waveform integrity during testing, a series inductor was installed at the output of the generator. This inductor compensated for capacitive loading from the test setup, stabilizing the rise time across all trials. Additionally, it served to dampen any high-frequency oscillations or voltage overshoot, ensuring that the square waveform stress delivered to the specimen was clean and repeatable.

Each TP sample was prepared according to the procedure detailed in Section II-A. It is acknowledged that partial discharge (PD) inception typically precedes dielectric breakdown in solid polymeric insulation, and most prior studies have therefore focused on PD measurement as the primary diagnostic tool. However, in the present work, conventional PD detection was not feasible due to the high-frequency nature of the applied voltage pulses. The state-of-the-art PD analyzer available (OMICRON MPD 800) has an upper frequency limit of approximately 10 kHz for the applied voltage, which is insufficient for the present study that extends up to 30 kHz. Moreover, under such conditions, the discharge current spectra overlap with the excitation harmonics, making reliable PD discrimination impractical. Therefore, insulation performance was characterized through voltage-ramp breakdown testing, providing breakdown-voltage measurements under the specified repetitive fast-edge excitation. Although partial discharge measurement was not feasible under the applied high-frequency square-wave excitation, several qualitative indicators consistent with discharge activity were observed during testing. Prior to complete breakdown, small transient perturbations and high-frequency oscillatory components were occasionally observed superimposed on the applied voltage waveform when monitored using the digital oscilloscope. In addition, audible acoustic activity was sometimes detected at higher stress levels immediately preceding breakdown. These observations are qualitatively consistent with localized discharge activity in inter-strand gaps. However, these signals were not separately quantified due to the absence of dedicated high-bandwidth current or acoustic sensors. This approach is consistent with methodologies adopted in pulsed-power insulation research, where the detection of individual PD events is often experimentally unfeasible.

The experimental program covered five frequencies between 1 and 30 kHz to compare the breakdown-voltage behavior of the two selected Litz-wire constructions. This range defines the scope of the present measurements and does not represent an established upper frequency limit for discharge initiation or breakdown. No measurements above 30 kHz were conducted; consequently, the observed trends should not be extrapolated to higher switching frequencies.

The experimental objective was to determine the breakdown voltage of Litz wire insulation across a range of switching frequencies. Once dielectric breakdown was detected during testing, the procedure was halted, the test setup fully discharged using a resistive discharge stick, and the sample replaced with a new one for the next test. A total of 25 TP samples were tested for each size of Litz wire, five samples for each selected frequency. The applied voltage amplitude was increased at a constant rate until breakdown. This controlled methodology enabled the acquisition of comparative data across frequencies for different Litz wire configurations. The entire test process was repeated at multiple switching frequencies to study the frequency-dependent breakdown behavior. The resulting data provided critical insight into the insulation performance of Litz wire under repetitive square-wave stress, simulating realistic voltage conditions imposed by WBG and UWBG-based power inverters.

## III. RESULTS AND ANALYSIS

### *A. Experimental Data*

Tables III and IV present the measured breakdown voltages for the 14 AWG and 16 AWG Litz-wire constructions, respectively, and Fig. 2 shows their mean values. Between 1 and 30 kHz, the mean breakdown voltage decreases from 620.6 to 554.0 V for the 14 AWG construction, approximately 10.7%, and from 631.4 to 505.4 V for the 16 AWG construction, approximately 20.0%. The 16 AWG mean decreases monotonically across the tested frequencies, whereas the 14 AWG mean increases from 620.6 V at 1 kHz to 655.4 V at 5 kHz before decreasing. These measurements quantify the different frequency sensitivities of the two tested constructions. With five specimens per condition, the modest increase at 5 kHz should be interpreted cautiously.

TABLE III
EXPERIMENTAL DATA FOR 14 AWG LITZ WIRE

| Sample No. | 1 | 2 | 3 | 4 | 5 | Mean |
|---|---|---|---|---|---|---|
| Frequency (kHz) | Breakdown Voltage (V) | | | | | |
| 1 | 567 | 586 | 634 | 644 | 672 | 620.6 |
| 5 | 644 | 644 | 644 | 663 | 682 | 655.4 |
| 10 | 586 | 586 | 591 | 600 | 624 | 597.4 |
| 20 | 548 | 548 | 552 | 562 | 586 | 560 |
| 30 | 514 | 552 | 552 | 557 | 591 | 554 |

TABLE IV
EXPERIMENTAL DATA FOR 16 AWG LITZ WIRE

| Sample No. | 1 | 2 | 3 | 4 | 5 | Mean |
|---|---|---|---|---|---|---|
| Frequency (kHz) | Breakdown Voltage (V) | | | | | |
| 1 | 600 | 609 | 619 | 638 | 691 | 631.4 |
| 5 | 542 | 566 | 590 | 614 | 619 | 586.2 |
| 10 | 552 | 552 | 537 | 585 | 614 | 568 |
| 20 | 542 | 547 | 552 | 561 | 566 | 553.6 |
| 30 | 489 | 499 | 508 | 513 | 518 | 505.4 |

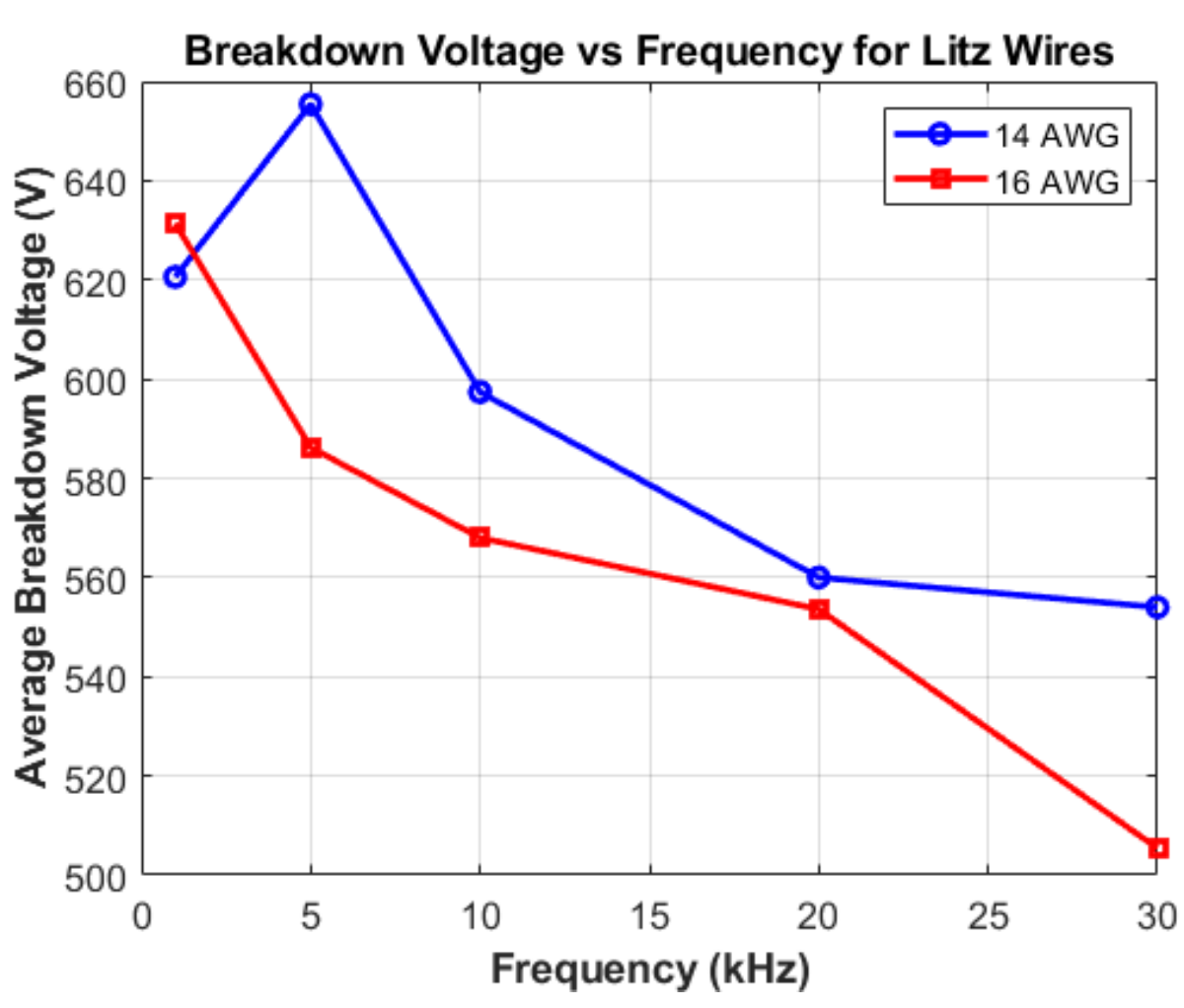

Fig. 2. Breakdown voltage vs. frequency characteristics of 14 AWG and 16 AWG Litz wire.

At 1 kHz, the mean breakdown voltage of the 14 AWG construction is slightly lower than that of the 16 AWG construction. At 5 kHz, the 14 AWG mean increases to 655.4 V. With five specimens per condition and no direct electric-field or discharge measurements, this increase cannot be assigned to a particular physical mechanism.

After this point, both wires experience a continuous decline in breakdown voltage. The 16 AWG wire shows a faster rate of degradation, and by 30 kHz, its breakdown voltage falls to about 505 V, compared with 554 V for the 14 AWG wire. Within the tested range, the 16 AWG construction exhibits a greater relative reduction in mean breakdown voltage than the 14 AWG construction; this comparison does not establish a general dependence on overall conductor diameter.

Charge accumulation, repetitive discharge activity, and dielectric heating are possible contributors to frequency-dependent breakdown behavior. Their individual contributions were not measured in this study. The greater relative reduction observed for the 16 AWG construction cannot be attributed to individual strand size because both constructions use 38 AWG strands. The comparison is specific to the tested constructions and excitation conditions; suitability for a particular application requires additional insulation qualification.

*B. Statistical Analysis*

The breakdown strength of insulating systems often exhibits significant statistical variability due to intrinsic material heterogeneity, microscopic defects, and the stochastic nature of partial discharges. Consequently, deterministic or simple arithmetic averaging techniques fail to capture the probabilistic nature of dielectric failure. Among several statistical tools available such as normal, log-normal, and exponential models the Weibull distribution has proven to be the most appropriate for characterizing breakdown voltage data. This suitability arises from its flexibility in representing both wide and narrow data spreads through its two parameters, as well as its direct physical interpretability: the shape parameter ($\beta$) quantifies the uniformity of breakdown behavior, while the scale parameter ($\alpha$) denotes the characteristic breakdown voltage corresponding to a 63.2 % probability of failure.

The Weibull distribution has become a widely used model in high-voltage insulation research because it captures the weakest-link mechanism inherent in dielectric failure. When a breakdown occurs at a microscopic imperfection, each potential failure site acts as an independent statistical unit. The cumulative failure probability $F(V)$ for such a population, follows:

$$F(V) = 1 - \exp\left[-\left(\frac{V}{\alpha}\right)^{\beta}\right] \tag{1}$$

where $V$ is the applied voltage. Linearization through a double-logarithmic transformation yield

$$\ln\big[-\ln\big(1 - F(V)\big)\big] = \beta \ln(V) - \beta \ln(\alpha) \tag{2}$$

which enables straightforward estimation of $\beta$ and $\alpha$ via linear regression. This transformed representation, often called a transformed Weibull probability plot, allows a linear fit of the breakdown data, as shown in Figs. 3 and 4.

As the present study involves a limited number of samples per frequency group, the parameter estimation follows the procedure prescribed in IEEE Std 930-2004 [23]. This standard recommends Ross's approximation to determine the cumulative failure probability for the $i^{\text{th}}$ ranked sample in a dataset of size

$n$ as follows:

$$F(i, n) = \frac{i - 0.44}{n + 0.25} \tag{3}$$

The fitting was performed with weighting factors to improve accuracy for small datasets. The fitted parameters describe the five-sample datasets and should be interpreted with consideration of statistical uncertainty. The extracted Weibull parameters for 14 AWG and 16 AWG Litz wire are summarized in Tables V and VI, respectively.

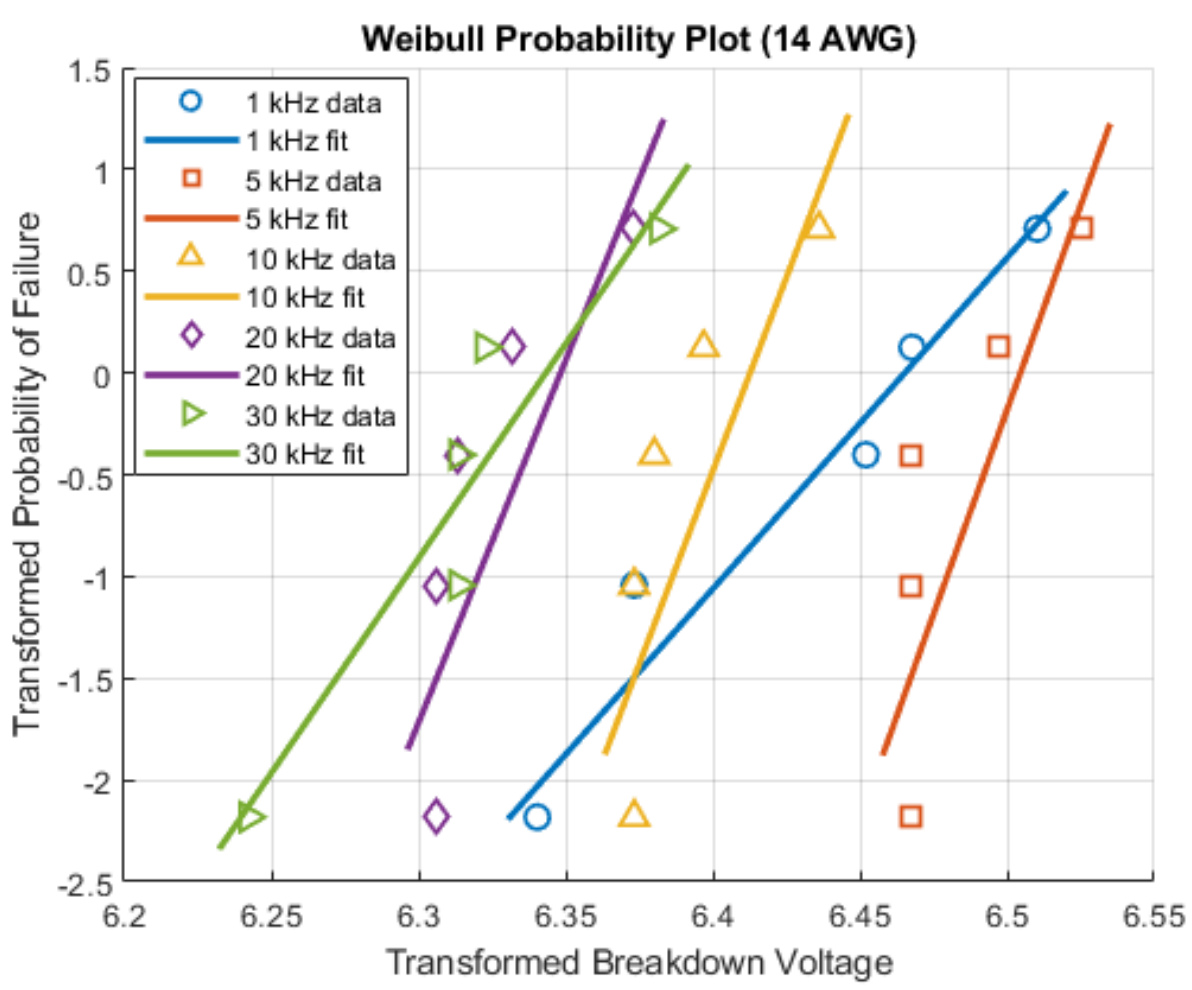

Fig. 3. Weibull probability plot for 14 AWG Litz wire.

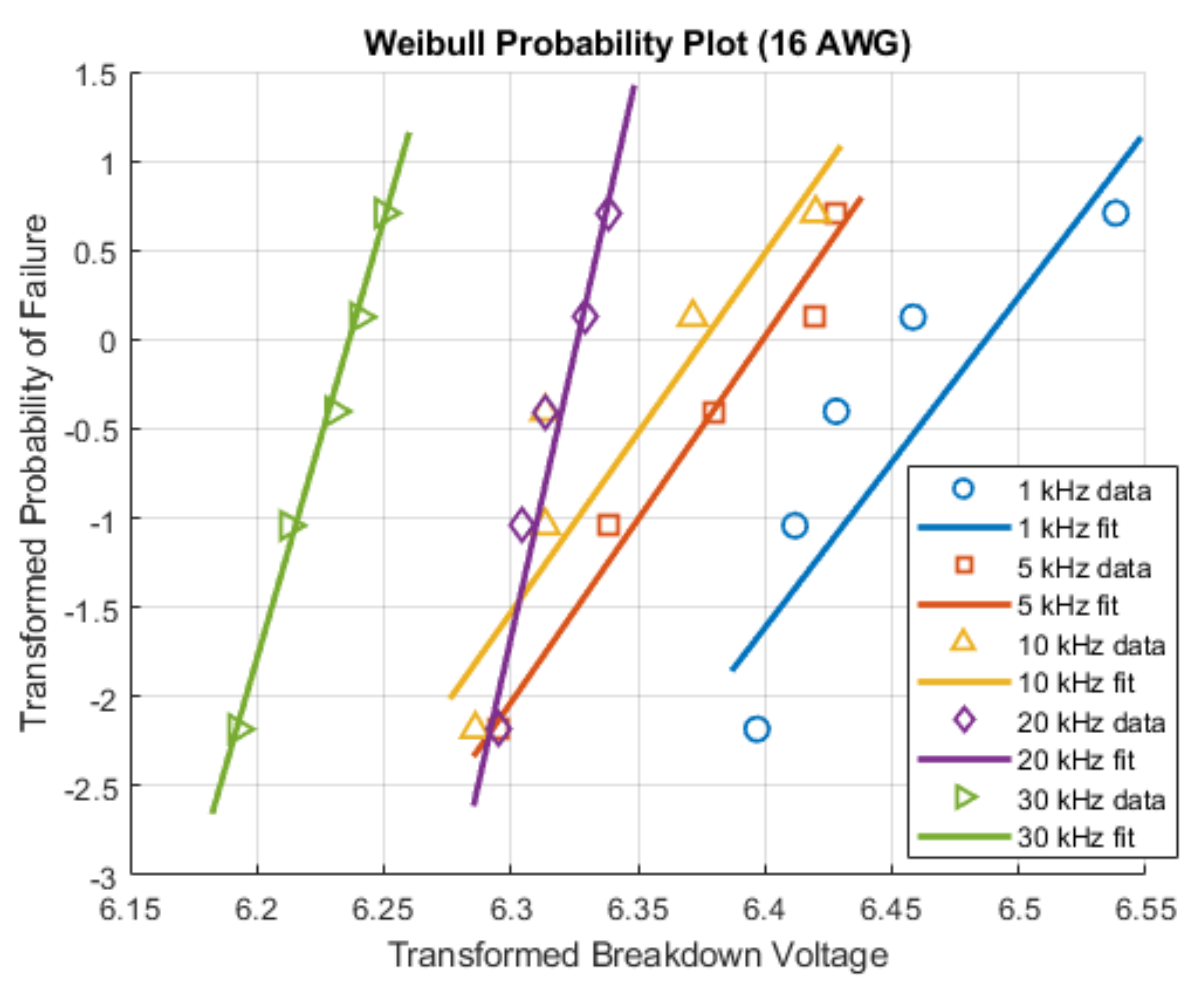

Fig. 4. Weibull probability plot for 16 AWG Litz wire.

TABLE V
SCALE AND SHAPE PARAMETERS FOR 14 AWG WIRE

| Frequency (kHz) | Scale Parameter ($\alpha$) | Shape Parameter ($\beta$) | Mean Breakdown Voltage (V) |
|---|---|---|---|
| 1 | 642.41 | 16.27 | 620.6 |
| 5 | 668.16 | 40.16 | 655.4 |
| 10 | 609.53 | 37.98 | 597.4 |
| 20 | 571.51 | 35.60 | 560 |
| 30 | 568.65 | 21.08 | 554 |

TABLE VI
SCALE AND SHAPE PARAMETERS FOR 16 AWG WIRE

| Frequency (kHz) | Scale Parameter ($\alpha$) | Shape Parameter ($\beta$) | Mean Breakdown Voltage (V) |
|---|---|---|---|
| 1 | 656.58 | 18.54 | 631.4 |
| 5 | 601.37 | 20.50 | 586.2 |
| 10 | 587.60 | 20.12 | 568 |
| 20 | 559.06 | 63.76 | 553.6 |
| 30 | 511 | 49.18 | 505.4 |

The fitted Weibull probability plots for the 14 AWG construction are shown in Fig. 3. The scale parameter α increases from 642.41 V at 1 kHz to 668.16 V at 5 kHz, then decreases to 568.65 V at 30 kHz. Thus, the measured characteristic breakdown voltage does not decline monotonically across the entire frequency range. The shape parameter β increases from 16.27 at 1 kHz to 40.16 at 5 kHz, then decreases to 21.08 at 30 kHz. These values describe differences in the dispersion of the measured breakdown voltages; they do not establish changes in the electric-field distribution or a specific degradation mechanism.

The mean breakdown voltage for 14 AWG wire increases from 620.6 V at 1 kHz to 655.4 V at 5 kHz (Tables III and V), and the corresponding Weibull scale parameter increases from 642.4 V to 668.2 V, while the Weibull shape parameter increases markedly from β = 16.3 to β = 40.2 (Table V). The simultaneous rise in both α and β indicates not only a higher characteristic breakdown strength but also substantially reduced dispersion at 5 kHz, which suggests a more statistically uniform breakdown process at that condition.

The increase in mean breakdown voltage between 1 and 5 kHz is approximately 5.6% and is accompanied by reduced dispersion in the five measured values. Given the small sample size, this difference should be interpreted cautiously. The measurements do not establish whether the increase reflects sampling variability or a reproducible physical effect. No direct measurements of charge relaxation, discharge energy, or temperature were obtained to determine its cause.

The Weibull analysis for the 16 AWG wire in Fig. 4 exhibits a similar trend but with lower characteristic breakdown voltages at 5–30 kHz. The characteristic voltage $\alpha$ declines from 656.58 V at 1 kHz to 511 V at 30 kHz, representing a greater reduction than in the 14 AWG case. The tested 16 AWG construction exhibits a greater relative reduction in characteristic breakdown voltage between 1 and 30 kHz than the 14 AWG construction. Although differences in bundle structure and contact geometry may contribute to this behavior, electric-field distributions and discharge inception were not measured, so the underlying cause cannot be established from the present data. The $\beta$ values for 16 AWG fluctuate over a broader range (18.5–63.8). The higher estimated $\beta$ at 20 kHz reflects tighter clustering of the measured breakdown voltages at that condition. With five specimens per frequency and no direct field or discharge measurements, this variation cannot be assigned to a specific physical mechanism. This comparison establishes different frequency sensitivities for the two tested constructions within the investigated range.

Between 1 and 30 kHz, the characteristic breakdown voltage decreases by approximately 11.5% for the 14 AWG construction and 22.2% for the 16 AWG construction. The 14 AWG scale parameter exceeds the 16 AWG value at 5–30 kHz, but is slightly lower at 1 kHz. The shape parameters vary nonmonotonically with frequency in both constructions, describing differences in the dispersion of the measured breakdown voltages. These results establish a comparison of the tested constructions under the specified conditions, but do not independently establish insulation lifetime or the physical mechanisms responsible for the observed differences.

Because each frequency group contains five samples, the estimated Weibull parameters should be interpreted with consideration of statistical uncertainty. The scale parameter $\alpha$ is generally more robust because it reflects the central tendency of the breakdown data, whereas the shape parameter $\beta$ is more sensitive to sample dispersion and individual early failures. For example, the five 14 AWG breakdown measurements at 5 kHz have a mean of 655.4 V and a sample standard deviation of approximately 17.0 V, giving an approximate two-sided 95% confidence interval for the mean of 634.3–676.5 V using the t-distribution with four degrees of freedom. This interval describes uncertainty in the estimated mean at 5 kHz; it does not, by itself, establish whether the difference between the 1 and 5 kHz means is statistically significant. Therefore, the Weibull parameters are used primarily to identify comparative trends across frequencies rather than exact population values.

### *C. Mechanistic Discussion*

Possible explanations for the measured frequency dependence include interactions among charge relaxation, repetitive discharge activity, and dielectric heating. Increasing frequency reduces the interval between voltage reversals and may affect residual charge, discharge development, and energy deposition. However, charge relaxation times, discharge activity, and specimen temperatures were not measured. The present data therefore do not establish which mechanism dominates or whether a transition from electrical to electrothermal breakdown occurs within 1–30 kHz. The following discussion presents possible mechanisms informed by the literature rather than mechanisms directly confirmed in the tested specimens.

The two constructions differ in overall conductor size, strand count, and bundle arrangement. These differences may influence contact geometry and local electric-field distribution, contributing to the observed breakdown behavior. However, both constructions use 38 AWG strands, and the effects of individual construction parameters were not isolated in this study.

To clarify the physical basis of residual charge accumulation and the resulting local electric-field enhancement at wire contact points under repetitive square-wave excitation, the following mechanism can be considered. In polymer insulation systems containing microvoids or interfacial gas gaps, discharge can initiate once the electric field within the void exceeds the gas breakdown strength. The presence of a void in a polymer dielectric causes local field distortion due to the permittivity difference between the gas cavity and the surrounding solid insulation. As discussed in [24], during a discharge event the generated charge carriers are driven toward the dielectric surface and a portion of these charges remains deposited on the surface after the discharge extinguishes. These residual charges modify the local boundary conditions of the electric field. Consequently, when the next voltage pulse is applied, the superposition of the applied electric field and the field generated by the residual surface charge produces local electric-field enhancement in the same region. This process promotes repeated discharge initiation at the same location. Such behavior creates a "memory effect," in which the previous discharge history influences subsequent discharge events and contributes to progressive stress concentration at localized interfaces.

In addition to surface charge deposition, charge transport and trapping within the polymer insulation also contribute to space-charge formation. As described in [25], electrons generated during microdischarge events can be deposited on the dielectric surface and subsequently diffuse into the polymer bulk. Once inside the material, these carriers may enter the conduction band and become trapped in localized energy states associated with structural defects or molecular irregularities. As the near-surface trap states become filled, additional charges accumulate in adjacent trapping levels, forming a persistent space-charge region near the interface. The presence of this trapped charge generates an internal electric field that modifies the local field distribution. Charge migration within the dielectric can then occur through thermally activated detrapping or hopping between trap states. In a twisted-pair Litz-wire configuration, the contact region between adjacent strands already represents a location of geometric field enhancement. The additional contribution of the internal space-charge field further increases the effective electric field in the inter-strand gap, increasing the probability of discharge initiation and accelerating insulation degradation.

Once local discharges occur repeatedly at these contact points, plasma-induced degradation mechanisms can develop. As discussed in [24], internal ionization in gas-filled voids leads to the formation of discharge channels in which energetic electrons and ions interact with the surrounding dielectric surface. A significant portion of the discharge energy is transferred to the polymer through electron and ion bombardment, which can produce localized heating, molecular bond scission, and chemical modification of the material. These processes gradually degrade the dielectric properties of the insulation and reduce its breakdown strength. Repeated discharge activity also promotes the formation of streamer-like channels within the gas cavity, further intensifying local stress.

Complementary mechanisms associated with plasma–polymer interaction are described in [25]. Transient microdischarges generate energetic electrons, ions, and reactive species that interact with the polymer surface and alter its electrical and interfacial properties. These interactions can modify the chemical structure of the polymer through oxidation, chain scission, or cross-linking reactions, leading to changes in surface morphology and electrical conductivity. Experimental observations reported in [25] also indicate that repeated plasma exposure produces measurable structural modifications on polymer surfaces. In enamel-coated Litz wires, similar processes can occur when microdischarges repeatedly strike the same inter-strand region. Localized plasma exposure can gradually erode the enamel layer, increase surface roughness, and introduce defect sites that further intensify electric-field concentration.

The combined effect of residual charge accumulation, space-charge formation, and plasma-induced surface modification creates a positive feedback mechanism. Residual charge enhances the local electric field, which promotes repeated discharge initiation. Repetitive discharge activity then modifies the polymer surface and introduces additional defects, further increasing the local electric-field concentration. Over time, this coupled electrostatic and plasma-assisted degradation process can significantly reduce the effective dielectric strength of the insulation and eventually lead to breakdown at the stressed wire contact interfaces.

It should be noted that the plasma-assisted degradation mechanism discussed above is inferred based on the observed breakdown behavior, localized damage patterns, and established discharge mechanisms reported in the literature. Direct microstructural characterization of the degraded enamel surface was not performed in the present study; therefore, the plasma-related interpretation should be considered a physically supported hypothesis rather than a directly confirmed observation.

### *D. Discussion and Implications*

The measurements demonstrate different breakdown-voltage responses for the two tested commercial Litz-wire constructions under the same excitation conditions. Because construction parameters were not varied independently, the results do not establish a causal dependence on overall conductor gauge, strand packing density, enamel thickness, or twist pitch. Charge accumulation, repetitive discharge activity, and dielectric heating may contribute to the observed frequency dependence, but their roles were not independently established. The measurements support assessing the tested constructions under the relevant repetitive voltage waveform rather than extrapolating their performance from power-frequency qualification alone. The reported breakdown voltages characterize the tested twisted-pair specimens; they do not establish intrinsic strand-level dielectric strength or complete winding withstand capability.

The Weibull parameters characterize the measured breakdown-voltage distributions under the specified voltage-ramp protocol. The scale parameter $\alpha$ corresponds to a cumulative breakdown probability of 63.2%, while the shape parameter $\beta$ describes the dispersion of breakdown voltages. A higher $\beta$ indicates tighter clustering of the measured values, but does not independently establish fewer insulation defects, manufacturing consistency, or longer service life. These parameters support comparison of the tested constructions; determining allowable operating voltages, insulation derating, or lifetime requires additional endurance measurements and application-specific qualification.

From an application standpoint, the measurements provide a quantitative comparison of two commercial Litz-wire constructions under identical excitation conditions. Because individual construction parameters were not varied independently, the results do not establish an optimum strand diameter, strand count, enamel thickness, or twist pitch. Selecting a construction for converter windings requires consideration of electrical losses, thermal conditions, mechanical constraints, and insulation performance under the intended operating waveform. Controlled comparisons and application-level qualification are needed to translate the present specimen measurements into winding-design recommendations.

The results are limited to the tested constructions, 200 ns rise time, and 1–30 kHz frequency range. They do not establish insulation performance at the higher switching frequencies used in many WBG converters. At higher frequencies, changes in pulse duration, charge relaxation, discharge activity, and dielectric heating may alter the breakdown response. Establishing these effects requires additional measurements and a validated physical model.

## IV. Conclusions

This study quantitatively compared the breakdown behavior of two commercial Litz-wire constructions under repetitive bipolar square-wave excitation with a 200 ns rise time over 1–30 kHz. The mean breakdown voltage decreased by approximately 10.7% for the 14 AWG construction and 20.0% for the 16 AWG construction between the lowest and highest tested frequencies. The 14 AWG response was not strictly monotonic, exhibiting a modest increase at 5 kHz before declining. Weighted Weibull analysis characterized the breakdown-voltage distributions, with shape parameters varying nonmonotonically across frequency. These findings provide construction-specific experimental data within the tested conditions.

The mechanistic discussion identified incomplete charge relaxation, space-charge accumulation, and localized heating as possible contributors to the observed frequency dependence. Their contributions and variation with frequency were not directly established by the present measurements.

Overall, the results provide construction-specific breakdown-voltage distributions for assessing the two tested Litz-wire constructions under repetitive fast-transient excitation. Establishing insulation derating, long-term reliability, or winding-design recommendations requires additional endurance testing and application-level qualification. Localized transient discharges and plasma-assisted degradation remain possible mechanisms requiring direct discharge diagnostics and post-mortem characterization for verification. Future work will extend these findings to include partial-discharge behavior and long-term endurance characterization under realistic converter operating conditions and will extend the study to impregnated and thermo-mechanically aged coil sections for enhanced application-level validation. Besides, post-mortem surface characterization, including optical microscopy and SEM analysis will be conducted, to further validate plasma-assisted degradation pathways under repetitive fast transient excitation.

Future work should extend the measurements above 30 kHz and combine breakdown testing with discharge diagnostics to evaluate whether the observed frequency dependence persists, saturates, or reverses.